\documentclass{icrc2009}

\usepackage{graphicx}   
\usepackage[font=footnotesize]{subfig} 
\usepackage{fixltx2e}

\newcommand{\shorttitle}[1]%
{\markboth{Poster shown at HEAD 2010, Big Island, Hawaii, March 1-4, 2010}{#1}}

\usepackage{url}
\usepackage{amsmath}
\usepackage{amsfonts}
\usepackage{amssymb}
\usepackage[numbers]{natbib}

\begin{document}
\title{Gamma-rays from Pulsar Wind Nebulae in Starburst galaxies}

\author{\IEEEauthorblockN{O.~Tibolla\IEEEauthorrefmark{1}\IEEEauthorrefmark{2}
			  K.~Mannheim\IEEEauthorrefmark{1},
			  D. Els\"asser\IEEEauthorrefmark{1}
}
                            \\
\IEEEauthorblockA{\IEEEauthorrefmark{1}Institut f\"ur Theoretische Physik und Astrophysik, Universit\"at W\"urzburg, D-97074 W\"urzburg, Germany}
\IEEEauthorblockA{\IEEEauthorrefmark{2}corresponding author: Omar.Tibolla@astro.uni-wuerzburg.de}
}
\shorttitle{PWNe in Starburst Galaxies, HEAD 2010}
\maketitle

\begin{abstract}

Recently, two nearby prominent starburst galaxies, M82 and NGC253, have been detected as point-like sources with gamma-ray telescopes at TeV energies $[1]$ $[2]$. It has been claimed that these
detections show that the cosmic ray intensity in the starburst galaxies is three orders of magnitude higher than in the Milky Way galaxy, assuming that the observed gamma rays arise due to pion
production of cosmic rays interacting with the ambient gas. The observed spectrum is flatter than the cosmic ray spectrum in the Milky Way galaxy, and this could be due to the much higher gas density
in the starburst galaxies $[3]$. The interpretation seems to be in line with the Ginzburg-model of the origin of cosmic rays according to which the cosmic rays are accelerated in the shells of supernova
remnants. As an immediate corollary it follows that the cosmic ray driven gamma ray luminosity should scale with the gas density and supernova rate. At lower energies, gamma-ray measurements
with the Fermi LAT instrument could provide support for this scaling $[4]$.
However, there are nagging doubts about the interpretation of the observations at very high energies. At the distance of the observed galaxies, point-like sources cannot be discriminated from diffuse
emission for an angular resolution of the order of 0.1$^{\circ}$.
Hence, the question about the contribution of unresolved point-like sources to the gamma-ray luminosity arises.

  \end{abstract}

 

\section{Cosmic-rays and pulsar wind nebulae (PWNe) in the Milky Way}

Towards an interpretation of the gamma ray emission from starburst galaxies, it is instructive to first take a look at the total TeV luminosity of the Milky Way galaxy.  The total luminosity can be obtained from adding the luminosity of gamma rays produced by cosmic rays interacting with the ISM and the luminosity of all point sources.  
The cosmic-ray driven luminosity due to pion production can be simply estimated to be

\begin{center}
 \begin{equation}  
\begin{split}
L_{\gamma , diff}(>1 \mathrm{GeV}) = \tau_{\pi} u_{CR} V \tau_{CR} =\\
= 3 \times 10^{38} \tau_{-2} u_1 V_{68} t_8^{-1} \mathrm{erg~s}^{-1}  		
\end{split} 
\end{equation}
\end{center}

where $\tau = n_{ISM} \sigma_{\pi} D = 0.01 \tau_{-2}$ denotes the optical depth for pion production, $u_1$ the energy density in units of 1 eV cm$^{-3}$, $V_{68}$ the galactic storage volume in units of $10^{68} \mathrm{cm}^3$, and $t_8$ the escape time in units of $10^8$ years . Extrapolation with $\Gamma = 2.7$ yields

\begin{equation}
L_{\gamma , diff}(>1 \mathrm{GeV})  = 2 \times 10^{36} \mathrm{erg~s}^{-1} 	
\end{equation}

This diffuse flux is difficult to measure with ground-based methods.  Space-born measurements up to energies of 100 GeV and the power law nature of the spectrum permit a reliable estimate of the observed TeV luminosity of the Milky Way galaxy.  One must bear in mind, however, that the observed diffuse flux is always contaminated with unresolved point sources.  Thus, the observed luminosity must be higher than the above estimate.
From the scan of the inner Galaxy performed with the H.E.S.S. array at TeV energies and the MILAGRO survey above 35 TeV, it is known that PWNe constitute the dominant source population at very high energies  $[5]$ $[6]$. There are good reasons to assume that the majority of so-far unidentified H.E.S.S. sources in the inner Galaxy are, in fact, ancient PWNe $[7]$.   Their dominance at TeV energies relates to the fact that they show rather hard spectra and have long life-times, corresponding to their synchrotron cooling times of the order of $t_{cool} = 130 (B/10  \mathrm{{\mu}G})^{-2} \mathrm{kyrs}$ at 1 TeV. The total TeV luminosity of the Galaxy due to the $2 \times 10^3 R_2$ expected PWNe (of which about $\sim$60 are known) should be

\begin{center}
\begin{equation}
\begin{split}
L_{PWN} (1-10 \mathrm{TeV}) = 1.6 \times 10^{37} (R/0.02 \mathrm{yr}^{-1}) \\
(L/10^{33.8} \mathrm{erg~s}^{-1}) (t_{cool}/130 \mathrm{kyrs}) 
\end{split}
\end{equation}
\end{center} 

Tacitly, we have assumed here that all core-collapse supernova do indeed produce a PWN (conservations of angular momentum and magnetic flux freezing).  Failed supernovae and black holes might not contribute to the census.  Averaging the sample of 33 PWNe from the compilation of [8] yields $log[<L>/\mathrm{erg~s}^{-1}] = 33.8$. Comparison with (2) shows that the PWN luminosity is a factor of about 20 higher than the diffuse cosmic-ray-induced gamma ray luminosity at TeV energies.   
As a matter of fact, the diffuse flux (including resolved and unresolved sources) at low galactic latitudes shows a change in slope due to the source populations with their flatter spectra at 10 GeV $[9]$.  The spectral slope of PWNe obtained from the compilation of $[8]$ is $<\Gamma>=2.3$, i.e. significantly flatter than the cosmic ray spectrum. Extrapolating the measured total intensity up to 1 TeV, the dominance of the source population component builds up to a factor of 10.  The intensity is then given by $1.4 \times 10^{-3} \mathrm{MeV~st}^{-1} \mathrm{cm}^{-2} \mathrm{s}^{-1}$, which corresponds to a luminosity of $2.5 \times 10^{37} \mathrm{erg s}^{-1}$, which is in fair agreement with the luminosity given by eq.(3).  Thus, the Milky Way galaxy seen at TeV energies is apparently dominated by pulsar wind nebulae.

\section{Total PWN luminosity of starburst galaxies and discusson}

 Estimating in the same way the PWN luminosity of starburst galaxies is straightforward. Inspection of eq. (3) reveals that the total luminosity is governed by the supernova rate only. Adopting values of $0.07 \pm 0.01 \mathrm{yr}^{-1}$ for NGC253 and of $0.11 \pm 0.05 \mathrm{yr}^{-1}$ for M82 yields the 1 TeV luminosities of $5.7 \times 10^{37} \mathrm{erg~s}^{-1}$ and $9 \times 10^{37} \mathrm{erg~s}^{-1}$, respectively. The expected PWN TeV luminosities are plotted in comparison with the 1 TeV luminosities observed with H.E.S.S. and VERITAS, including an upper limit for Arp220 obtained with the MAGIC Telescope $[10]$. The TeV luminosities have been computed from the observed fluxes by using the average spectral index $<\Gamma>=2.3$ of the known TeV-emitting PWNe. Results are $L_{NGC253} = (3.6 \pm 2.6) \times 10^{38} \mathrm{erg~s}^{-1}$ adopting a distance of 3.337 Mpc, and $L_{M82} = (1.8 \pm 0.9) \times 10^{39} \mathrm{erg~s}^{-1}$ for a distance of 4.545 Mpc. For the Milky Way galaxy, we extrapolate the total (diffuse+source) luminosity from the low galactic latitude observations by Fermi up to 1 TeV (from $[9]$).
PWNe naturally account for the observed VHE gamma ray emission from starburst galaxies. At TeV energies, the total flux from the population of PWNe readily surmounts the diffuse flux due to cosmic rays. In the Milky Way galaxy, the dominance of PWNe sets in at about 10 GeV, as can be seen from the analysis of the Fermi data on the Galactic foreground emission. This is also in agreement with the H.E.S.S. scan of the inner Galaxy at TeV energies and the MILAGRO survey at energies above 35 TeV.
Since the PWNe are obviously the strongest source population on the VHE sky, they should also be reconsidered as sources of cosmic rays. The cosmic ray luminosity due to shell-type supernova remnants is given by $L_{CR} = 6 \times 10^{40} R_2 E_{51} \epsilon_{0.1} \mathrm{erg~s}^{-1}$ where $R_2$ denotes the supernova rate in units of 2 per century, $E_{51}$ the kinetic energy in units of $10^{51}$ ergs, and $\epsilon_{0.1}$ for the acceleration efficiency in units of 10\%. Rather high efficiencies for the neutrino-to-kinetic and kinetic-to-cosmic ray energy conversions are required. Tapping the rotational energy of newborn neutron stars is less troubled with these bottlenecks, the highly magnetized pulsar equatorial wind providing the most efficient imaginable machine.
The recent measurement of an excess in the cosmic positron fraction by the PAMELA collaboration may demonstrate that the particles accelerated at pulsar wind shocks in the Local Bubble are released into the ISM after disruption of the pulsar wind bubble in the absence of adiabatic losses. Protons and ions, swept into the plerion along by the reverse shock, can be accelerated efficiently at the ultrarelativistic pulsar wind shocks $[11]$ $[12]$.
Searching for the sources of cosmic rays, special attention should thus be given to PWNe in their late phases. Current high-energy neutrino measurements do not give a hint yet $[13]$, however, a stacked source analysis of PWNe could enhance the sensitivity to their putative neutrino flux.

\section{References}
$[1]$ H.E.S.S. collaboration (2009), \emph{Science}, {\bf 326}, 1080  

$[2]$ Acciari et al. (VERITAS) (2009), \emph{Nature}, {\bf 462}, 770  

$[3]$ Persic et al. (2008), \emph{A\&A}, {\bf 486}, 143            

$[4]$ Abdo et al. (Fermi LAT) (2009), \emph{ApJ}, {\bf 709}, L152   

$[5]$ Aharonian et al. (H.E.S.S.) (2006), \emph{ApJ}, {\bf 636}, 777 

$[6]$ Abdo et al. (MILAGRO) (2009), \emph{ApJ}, {\bf 700}, L127   
  
$[7]$ de Jager et al. (2009),  arXiv:0906.2644

$[8]$ Kargalstev \& Pavlov (2009), arXiv:1002.0885 

$[9]$ Strong et al. (Fermi LAT)(2009), Fermi Symposium 2009 

$[10]$ Albert et al. (MAGIC) 2007), \emph{ApJ}, {\bf 658}, 245 

$[11]$ Atoyan \& Aharonian (1996), \emph{MNRAS}, {\bf 278}, 525

$[12]$ Bednarek \& Bartosik (2005), \emph{Journal of Phys. G Nucl and Part. Phys}, {\bf 31}, 1465 

$[13]$ Abbasi et al. (AMANDA) (2009), \emph{PhRvD}, {\bf 79}, 2001

\end{document}